\documentclass[prb,amsmath]{revtex4}
\usepackage{graphicx}
\usepackage{bm}

\begin{document}
\title{Photoconductiviy of 2D Rashba system in the perpendicular AC magnetic field}
\author{I. I. Lyapilin}
\email{Lyapilin@imp.uran.ru}
\affiliation{Institute of Metal Physics, UD of RAS\\
18, S.Kovalevskaya St. GSP-170 Ekaterinburg 620219 Russia}
\author{A. E. Patrakov}
\affiliation{Institute of Metal Physics, UD of RAS\\
18, S.Kovalevskaya St. GSP-170 Ekaterinburg 620219 Russia}

\begin{abstract}
The response of a 2D electron system to a DC measurement electric field has
been investigated in the case when the system is driven out of the
equilibrium by the magnetic ultra-high frequency field that leads to
combined transitions involving the spin-orbit interaction.
It has been shown that the method of non-equilibrium statistical operator
in conjunction ith the method of canonical transformations
allows one to build a theory of linear response of a non-equilibrium
2D electron gas to a weak ``measurement'' DC electric field.
The proposed theory predicts that such perturbation of the electron system
with high ($\sim 10^7$~cm$^2$/Vs) mobility leads to a new type
of 2D electron gas conductivity oscillations
controlled by the ratio of the radiation frequency to the cyclotron frequency.
\end{abstract}
\maketitle
\section{Introduction}
In 2D electron systems with high ($\sim 10^7$~cm$^2$/Vs) electron mobility,
the magnetoresistance exhibits strong oscillations in ``pre-Shubnikov''
range of magnetic fields under microwave irradiation \cite{Zudov03,Mani02}.
These oscillations are caused by electron transitions between Landau levels
due to the electric component of the microwave radiation.

Together with the oscillations of the diagonal components of the
conductivity tensor,
``beats'' have been found experimentally \cite{Mani02} in the interval of
more weak magnetic fields.
Such beats are usually related with the manifestations of the interaction
between kinetic and spin degrees of freedom of the conductivity electrons.
Such interaction is the spin-orbit interaction (SOI) that is known to be the
origin of numerous effects in transport phenomena observed in such systems
\cite{Hammar-00,Das,Levitov}, etc.
SOI also leads to the possibility of electron transitions between Landau
levels in the magnetic field at the combined resonance frequencies
\cite{Rash}, thus transitions being possible both in antinodes of electric
and magnetic fields \cite{Kalash}.
Finally, operation of a spin transistor (schemes of which have been
considered in \cite{zutic}) is based upon spin degrees of freedom.
All of the above has determined elevated interest to investigations of the
SOI in 2D semiconductor structures.

For the purpose of studying the SOI, it appears interesting to investigate a
model in which the role of the SOI should manifest itself the strongest.
Since the SOI depends upon both translational and spin degrees of freedom,
then it is a channel over which energy (both electric and magnetic) can be
absorbed from the ultra-high frequency field, thus causing transitions
between Landau levels.
Because of that, it is interesting to investigate the response of a
non-equilibrium electron system to a DC weak (``measurement'') electric
field for the case when the initial non-equilibrium state is created with a
high-frequency AC magnetic field that leads to combined transitions.
The question is how this perturbation affects transport coefficients, in
particular, the conductivity tensor.

\section{Effective Hamiltonian}
The discussed model includes the contributions form Landau quantization and
(in the long-wavelength limit) from the microwave radiation.
We consider impurity centers for the role of scatterers, treating the
scattering process perturbatively.

The Hamiltonian of the system under consideration consists of the
kinetic energy $H_k$, Zeeman energy $H_s$ in the magnetic field $\bm H =
(0,0,H)$, the spin-orbit interaction $H_{ks}$, interactions of electrons
with AC magnetic and DC electric fields and with impurities, and the
Hamiltonian of the impurities themselves:
\begin{equation}\label{1}
\mathcal H(t) = H_k+H_s+H_{ks}+H_{eh}(t)+H^0_{ef}+H_v+H_{ev},
\end{equation}
$$
 H_k=\sum_j\frac{(\bm p_j-(e/c)\bm A (x_j))^2}{2m},
$$
$$
 H_s=\hbar\omega_s\sum_j S_j^z,\quad \hbar\omega_s=g\mu_0H.
$$
$S_i^\alpha$ and $p_i^\alpha$ are operators of the components of
the spin and kinetic momentum of the $i$th electron, where
$[p_i^\alpha,p_j^\beta] = -i\delta_{ij} m
\hbar\omega_c\varepsilon_{\alpha\beta z}$, $\omega_c=|e|H/mc$ is
the cyclotron frequency, and $\mu_0$ is Bohr magneton.

In this paper, we limit our consideration to the case when AC and DC
magnetic fields are parallel to each other: $\bm H(t) = (0, 0, H^z(t))$.
In this case, the Hamiltonian of the interaction of electrons with the
AC magnetic field has the form:
\begin{equation}\label{002}
 H_{eh}(t) = g\mu_0  H^z(t) \sum_j  S_j^z.
\end{equation}

We assume the specific form of the SOI term, namely, Rashba interaction,
which is non-zero even in the linear order in momentum:
\begin{equation}\label{7}
H_{ks}(p) =
\alpha \varepsilon_{zik} \sum_j S^i_j p^k_j =
\frac{i\alpha}{2} \sum_j (S^+_j p^-_j - S^-_j p^+_j),
\end{equation}
$$
S^\pm = S^x \pm i S^y, \quad p^\pm = p^x \pm i p^y.
$$
Here $\alpha$ is the constant characterizing the SOI, $\varepsilon$ is the
fully-antisymmetric Levi---Chivita tensor.

The spin-orbit interaction leads to correlation of spatial and spin motion
of electrons, thus, the translational and spin-related subsystems are not
well-defined.
Since the SOI is in some sense small, then one can perform a
momentum-dependent canonical transformation that decouples kinetic and spin
degrees of freedom.
All other terms in the Hamiltonian, describing the interaction
of electrons with the lattice and external fields also undergo the
transformation.
In this case, the effective interaction of electrons in the system with
external fields appears, which leads to resonant absorption of the field
energy not only at the frequency of the paramagnetic resonance $\omega_s$
or cyclotron resonance $\omega_c$, but also at their linear
combinations.
The gauge-invariant theory describing such transitions has been developed in
\cite{VPKalash}.

Assuming the SOI to be small, we perform the canonical transformation of the
Hamiltonian.
Up to the terms linear in $T(t)$, we have:
\begin{equation}\label{5}
\tilde {\mathcal H} = e^{-i T(p)}\mathcal H e^{i T(p)}\approx
\mathcal H -i [T(p), \mathcal H].
\end{equation}
The operator of the canonical transformation $T(p)$ has to be determined
from the requirement that, after the transformation, the $k$ and $s$
subsystems become independent. This requirement, as one can easily see,
is satisfied if one puts
\begin{equation}\label{8}
T(p)=-\frac{\alpha}{2\hbar(\omega_c-\omega_s)}
\sum_j(S^+_jp^-_j-S^-_jp^+_j).
\end{equation}

One can write the transformed Hamiltonian in the following form:
\begin{equation}\label{9}
\mathcal{\tilde{H}}(t)=H_0 + H^0_{ef}
+H_{eh}(t)+[T(p),H_{eh}(t)+H^0_{ef}+{H}_{ev}],
\end{equation}
$$
H_0 = H_k + H_s + H_v + H_{ev}.
$$
The effective interaction of the electrons and the AC magnetic field
(responsible for combined transitions) can be found using the explicit
expression for the operator $T(t)$:
\begin{equation}\label{10}
H_{eh,1}(t) = -i [T(p),H_{eh}(t)] =
\frac{i\alpha\omega_{1s}}{2(\omega_c - \omega_s)}
(T^{+-} - T^{-+})\cos \omega t,
\end{equation}
$$
T^{\alpha \beta}=\sum_i S^\alpha_i p^\beta_i,
$$
where $\omega_{1s}=g e H_1/(2 m_0 c)$,
$H_1$ is the intensity of the
linearly polarized magnetic field, oscillating with the frequency
$\omega$ according to the cosine law.

It follows from Eq. (\ref{10}) that the effective
interaction $H_{eh,1}(t)$ leads to combined transitions at frequency
$\omega_c - \omega_s$, while
the interaction of the spin degrees of freedom of the conductivity electrons
with the AC magnetic field $H_{eh}(t)$ leads to resonant transitions at the
frequency $\omega_s$.
Since, for our further calculations, the response of the non-equilibrium
system to the measurement electric field is interesting, in which the
contribution from the translational degrees of freedom dominates, we will
restrict our consideration to the effective interaction solely.

During the calculation of the non-equilibrium response of the electron
system to the measurement electric field, there are technical
difficulties cased by the dependence of the effective interaction
$H_{eh,1}(t)$ upon time.
One can avoid them by transferring the time dependence to the impurity
subsystem.
This can be done by means of a new canonical transformation.
The explicit form of the canonical transformation $W_2(t)$ is determined
from the requirement that it excludes $H_{eh,1}(t)$ from the effective
Hamiltonian of a system without impurities:
\begin{equation}\label{eq:UC2}
W_2^\dagger(t) (-i\hbar \frac{\partial}{\partial t} +H_k + H_s +
H_{eh,1}(t))W_2(t) = -i\hbar \frac{\partial}{\partial t} +H_k + H_s.
\end{equation}
The operator $W_2(t)$ is expressed in the following form:
\begin{equation}\label{w2}
    W_2(t) = \exp(i T_2(t)).
\end{equation}
The operator $T_2(t)$ is searched for in the form:
\begin{equation}\label{eq:UC2p}
T_2(t)=\eta^+(t) T^{+-} + \eta^-(t) T^{-+},
\end{equation}
where one has to determine the parameters $\eta^\pm(t)$.
In the linear approximation in the constant of the spin-orbit
interaction, we have:
\begin{equation}
\eta^\pm(t)=\frac{\alpha \omega_{1s}(
(\omega_c - \omega_s)\cos \omega t \pm i \omega \sin \omega t)}
{2 \hbar (\omega_c - \omega_s) ((\omega_c - \omega_s)^2 - \omega^2)}
\end{equation}

As a result of the canonical transformation $W_2(t)$,
renormalization of the electron-impurity interaction happens.
In the case of elastic scattering, for
obtaining the renormalized Hamiltonian of the electron-impurity
interaction, it is sufficient to calculate
$W_2^\dagger(t) \exp(i \bm q \bm r_j) W_2(t)$.
In the linear approximation in the constant of the spin-orbit
interaction $\alpha$, we obtain:
\begin{equation}
W_2^\dagger(t) \exp(i \bm q \bm r_j) W_2(t) = \exp(i \bm q \bm r_j)
(1 - i \hbar(\eta^+(t)S_j^+ q^- + \eta^-(t) S_j^- q^+))
\end{equation}
Using the explicit form of $\eta^\pm(t)$, we have:
\begin{multline}\label{ei}
W_2^\dagger(t) \exp(i \bm q \bm r_j) W_2(t) = \exp(i \bm q \bm r_j)
-i \frac{\alpha \omega_{1s}}{2 (\omega_c - \omega_s)
((\omega_c - \omega_s)^2 - \omega^2)} \\\times
\bigg(\Big((\omega_c - \omega_s) \cos \omega t + i \omega \sin \omega t
\Big) S_j^+ q^- +
\Big((\omega_c - \omega_s) \cos \omega t - i \omega \sin \omega t
\Big) S_j^- q^+ \bigg) \exp(i \bm q \bm r_j)
\end{multline}
The speed of the electron momentum change is:
\begin{equation}
\dot p^\pm_{(\tilde v)} = \dot p^\pm_{(v)} - \sum\limits_{\bm q j}
\sum\limits_{k,l=\pm 1} V(q) \rho(q) \frac{\alpha \omega_{1s}}{2
(\omega_c - \omega_s)} q^\pm \exp(i \bm q \bm r_j) \frac{S_j^k
q^{-k} e^{i l \omega t}}{\omega_c - \omega_s - k l \omega},
\end{equation}
where
\begin{equation}
S_j^k = \left\{
\begin{array}{l}
S_j^+, k=+1 \\
S_j^-, k=-1
\end{array}
\right.,\qquad
q^k = \left\{
\begin{array}{l}
q^+, k=+1 \\
q^-, k=-1
\end{array}
\right.
\end{equation}

As one can see from (\ref{ei}), the renormalized electron-impurity
interaction Hamiltonian acquired time dependence.
In such canonically transformed system, impurities act as a coherent
oscillating field that leads to resonant transitions.

\section{Momentum Relaxation Rate}
We assume that the initial non-equilibrium state of the system under
consideration is created by the ultra-high frequency magnetic field and
can be described with the distribution $\bar{\rho}(t)$.
If some
additional perturbation acts upon the system, then a new non-equilibrium
state is formed in the system, that requires an extended set of basis
operators for its description.
The new non-equilibrium distribution is described with the operator
$\rho(t,0)$.
The task is to find the response of a non-equilibrium system to a weak
measurement field.
Using the technique of calculation the non-equilibrium statistical
operator \cite{VPK}, we have for the momentum relaxation rate:
\begin{equation}\label{t1}
\frac{1}{\tau} = \frac{1}{2 m n T} \operatorname{Re}
\frac{1}{i\hbar} \int_{-\infty}^0 dt_1 e^{(\varepsilon -
\omega_1)t_1}  dt_2 e^{\varepsilon t_2} \int_0^1 d\lambda A(\lambda,
t_1 + t_2),
\end{equation}
\begin{equation}
A(\lambda, t_1 + t_2) = \operatorname{Sp}\{ \dot p^+_{(\tilde v)}(t)
e^{i L_0 (t_1 + t_2)}\rho_q^{\lambda}[\dot p^-_{(\tilde
v)}(t+t_1+t_2), H_k + H_s] \rho_q^{ 1-\lambda}.
\end{equation}
Here $T$ is the common temperature of the kinetic and spin subsystem,
that can be introduced when one neglects the ``heating'' effects,
$\rho_q(t)$ is the quasiequilibrium statistical operator.

Now we expand the formula (\ref{t1}) using the explicit expression
for the renormalized electron-impurity interaction
$\tilde{H}_{ev}$.
Inserting the explicit expression for the electron-impurity
interaction and averaging over the system of scatterers, we
obtain:
\begin{multline}\label{t3}
A(\lambda, t_1 + t_2) = \sum\limits_{\bm q j j'} \sum\limits_{k,l=\pm 1}
|V(\bm q)|^2 N_i
\frac{\alpha^2 \omega_{1s}^2}
{4(\omega_c - \omega_s)^2 (\omega_c - \omega_s - k l \omega)^2} q^4\\\times
e^{-i l \omega (t_1 + t_2)}
\operatorname{Sp}\{
S_j^k \exp(i \bm q \bm r_j)
e^{i L_0 (t_1 + t_2)}
\rho_q^{\lambda}
[ S_{j'}^{-k} \exp(-i \bm q \bm r_{j'}), H_k + H_s]
\rho_q^{1-\lambda}
\}
\end{multline}

Here $N_i$ is the impurity concentration. Rewriting this
expression using secondary quantization and averaging the Fermi
operators using Wick's theorem, we have:
\begin{multline}
A(\lambda, t_1 + t_2) = \sum\limits_{\bm q \nu \mu}\sum\limits_{k,l=\pm
1}
|V(\bm q)|^2 N_i
\frac{\alpha^2 \omega_{1s}^2}
{16(\omega_c - \omega_s)^2 (\omega_c - \omega_s - k l \omega)^2} q^4\\\times
e^{-i l \omega (t_1 + t_2)}
(\varepsilon_{\mu'} - \varepsilon_{\mu})
|(2 S^k e^{i \bm q \bm r})_{\nu \mu}|^2
f_\nu (1 - f_\mu)
e^{-\frac{i}{\hbar}(t_1 + t_2)(\varepsilon_\mu - \varepsilon_{\nu})}
e^{-\beta_e(\varepsilon_\mu - \varepsilon_{\nu})\lambda},
\end{multline}
where $f$ is the Fermi---Dirac distribution. $\nu$ and $\mu$ are the
electron states in the DC magnetic field, they are characterized by the
Landau level number $n$, $x$ projection of the wave wector $k^x$, and
$z$ projection of the spin $s^z$.

Then we should integrate over $\lambda$, $t_1$ and $t_2$, and take the limit
$\omega_1 \to 0$, $\varepsilon\to +0$ (because we are interested in the
zero-frequncy response). For the correction to the momentum relaxation
rate caused by the microwave radiation, we obtain:
\begin{multline}\label{t5}
\int_{-\infty}^0 dt_1 e^{(\varepsilon - i \omega_1) t_1}
 dt_2 e^{\varepsilon t_2} \int_0^1
d\lambda A(\lambda, t_1 + t_2) =\sum\limits_{\bm q \nu \mu l}
|V(\bm q)|^2 N_i J_l^2(|K_q|) q^2|(2 S^z e^{i \bm q \bm r})_{\nu
\mu}|^2 (f_\nu - f_\mu)\\\times
\frac{1}{\varepsilon - i l
\omega - (i/\hbar)(\varepsilon_\mu - \varepsilon_\nu)}
\frac{1}{\varepsilon - i\omega_1 - i l \omega - (i/\hbar)(\varepsilon_\mu
- \varepsilon_\nu)}
\end{multline}
In the $\varepsilon \to 0$ limit, we obtain:
\begin{equation}\label{t7}
\Delta(\frac{1}{\tau}) = -\sum\limits_{\bm q \nu \mu k l}
|V(\bm q)|^2 N_i
\frac{\pi \hbar \alpha^2 \omega_{1s}^2}
{8 m n (\omega_c - \omega_s)^2 (\omega_c - \omega_s - k l \omega)^2} q^4
|(S^k e^{i \bm q \bm r})_{\nu \mu}|^2
(f_\nu - f_\mu)
\frac{\partial}{\partial \varepsilon_\mu} \delta(l \hbar \omega +
\varepsilon_\mu - \varepsilon_\nu)
\end{equation}
The equation (\ref{t7}) contains a singularity in its right hand side, which
is removed, as usual, due to broadening of the Landau levels by scattering
electrons on impurities:
\begin{equation}\label{t8}
\delta(\mathcal E - \varepsilon_\mu) \to D_\mu(\mathcal E)=
\frac{\sqrt{\pi/2}}{\Gamma} \exp\left(-\frac{(\mathcal E -
\varepsilon_\mu)}{2 \Gamma^2}\right).
\end{equation}
The Landau level width $\Gamma$ can be expressed via the electron mobility
$\mu$ in zero magnetic field:
\begin{equation}\label{t9}
\Gamma = \hbar\sqrt\frac{2 \gamma_n \omega_c}{\pi \tau_{\textrm{tr}}}, \qquad
\tau_{\textrm{tr}}=\frac{m \mu}{|e|}
\end{equation}

Integrating over the energy for the case $T > \Gamma$, we have:
\begin{equation}\label{t10}
\int d\mathcal E \frac{\partial}{\partial \mathcal E}
D_\nu(\mathcal E \pm \hbar\omega) D_\mu(\mathcal E) \\
=-\frac{\pi^{3/2}(\varepsilon_\mu - \varepsilon_\nu \pm
\hbar\omega)}{4\Gamma^3} \exp\left(-\frac{(\varepsilon_\mu -
\varepsilon_\nu \pm \hbar\omega)^2} {4 \Gamma^2}\right)
\end{equation}

Calculating the matrix element in (\ref{t7}) on the wave functions
\begin{equation}\label{t11}
\psi_{\nu}\equiv\psi_{n k^x S^z} = \frac{1}{\sqrt{2^n n! \pi^{1/2}
\ell}} \exp(i k^x x)
\exp\left(-\frac{(y - y_0)^2}{2 \ell^2}\right) H_n(\frac{y -
y_0}{\ell})\chi_{S^z},
\end{equation}
we obtain:
\begin{multline}
|\langle n_\nu k^x_\nu S^z_\nu | S^k \exp(i \bm q \bm r) | n_\mu
k^x_\mu S^z_\mu \rangle|^2 = \delta_{S^z_\nu + k, S^z_\mu}
\delta_{k^x_\nu, q_x + k^x_\mu} \exp\left(-\frac{\ell^2
q^2}{2}\right) \\\times
\frac{(\min(n_\nu,
n_\mu))!}{4(\max(n_\nu, n_\mu))!} \left(\frac{\ell^2
q^2}{2}\right)^{|n_\nu - n_\mu|} \left(L_{\min(n_\nu,
n_\mu)}^{|n_\nu - n_\mu|}\left( \frac{\ell^2
q^2}{2}\right)\right)^2.
\end{multline}
Here $y_0=\ell^2 k^x$ is the cyclotron orbit center coordinate, $\ell$
is the magnetic length, $H_n(x)$ denotes Hermite polynomials, and
$\chi_{S^z}$ is the eigenfunction of the $z$ spin projection.

Finally, integrating over $\bm q$, we have:
\begin{multline}
\int\limits_0^\infty d(q^2) q^4 \exp\left(-\frac{\ell^2 q^2}{2}\right)
\left(-\frac{\ell^2 q^2}{2}\right)^{|n_\nu - n_\mu|}
\left(L_{\min(n_\nu, n_\mu)}^{|n_\nu - n_\mu|}(\frac{\ell^2
q^2}{2})\right)^2 \\=
\frac{8}{\ell^6}\frac{(\max(n_\nu, n_\mu))!}{(\min(n_\nu, n_\mu))!}
(n_\nu^2 + n_\mu^2 + 3 (n_\nu + n_\mu) + 4 n_\nu n_\mu + 2)
\end{multline}

Thus, for the case of point scatterers, where $V(q)$ does not depend on
$q$, the radiation-induced correction to the inverse relaxation time is:
\begin{multline}\label{t16}
\Delta(\frac{1}{\tau}) = \sum\limits_{n_\nu n_\mu k l} |V(q)|^2 N_i
\frac{\pi^{1/2}\hbar\alpha^2 \omega_{1s}^2}
{32 m n \ell^8 \Gamma^3 (\omega_c - \omega_s)^2 (\omega_c - \omega_s - k
l \omega)^2}\\\times
(n_\nu^2 + n_\mu^2 + 3 (n_\nu + n_\mu) + 4 n_\nu n_\mu + 2)
(f(\varepsilon_\nu)-f(\varepsilon_\mu))\\\times
((n_\mu - n_\nu) \hbar \omega_c + k \hbar\omega_s + l \hbar\omega)
\exp\left(-\frac{(
(n_\mu - n_\nu) \hbar \omega_c + k \hbar\omega_s + l \hbar\omega
)^2}
{4\Gamma^2}\right)
\end{multline}

\section{Numerical Analysis}
Using the expression for the momentum relaxation rate, one can also write
the formula for the diagonal components of the conductivity tensor
$\sigma_{x x}=(n m^{-1} e^2 \tau)/(\omega_c^2\tau^2 + 1)$.
Numerical calculations have been carried
out with the following parameters: $m = 0.067\ m_0$ ($m_0$ is the
free electron mass), the Fermi energy is $\mathcal{E}_F = 10$~meV,
the mobility of the 2D electrons varies as $\mu\approx 0.9 -
1.5\times 10^7$~cm$^2$/Vs, the electron density $n=3\times
10^{11}$~cm$^{-2}$. The microwave radiation frequency is $f=
50$~GHz, the temperature is $T\approx 2.4$~K. The magnetic field
varied as 0.02~--~0.3~T.

The dependence of the 2D electron gas photoconductivity on the
$\omega/\omega_c$ ration is presented in Fig.~1.
\begin{figure}[t]
\center\includegraphics[width=10cm]{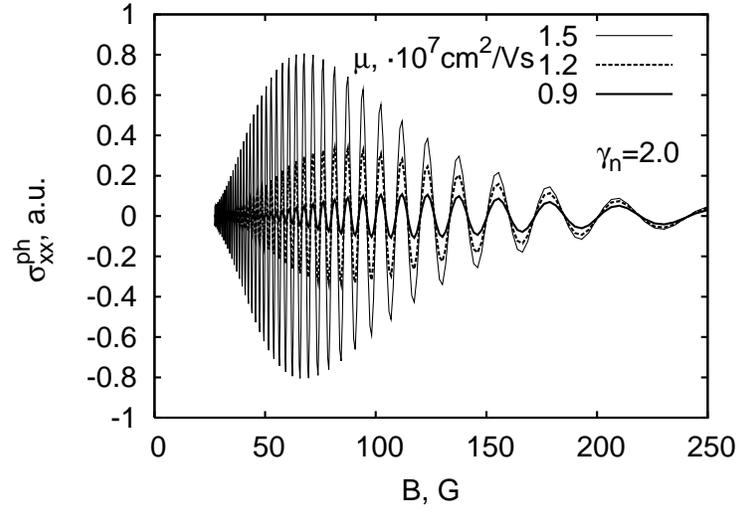}
\caption{Photoconductivity of the 2D electron gas vs of the
magnetic field induction for different values of electron
mobility. The radiation frequency is 50~GHz and
$\gamma=2$.}\label{fig:Lmuhmu}
\end{figure}
One can see that the dependence of electron mobility upon the
magnetic field has the oscillating character.

\section{Conclusion}
The response of a non-equilibrium electron system to the DC electric
measurement field has been studied for the case when the initial
non-equilibrium state of the system is created by an ultra-high frequency
magnetic field that leads to combined transitions.
Within the proposed theory, it has been shown that such
perturbation of the electron system essentially
influences the transport coefficients and leads to the oscillations of the
diagonal components of the conductivity tensor.
The discussed effect is analogous to the phenomenon observed in
GaAs/AlGaAs heterostructures with ultra-high electron mobility \cite{Mani02}.
However, unlike that phenomenon, the manifestation of the oscillatory
pattern is dictated by the spin-orbit interaction existing in the crystals
under consideration.

\end{document}